\documentclass[conference]{IEEEtran}
\IEEEoverridecommandlockouts
\usepackage{cite}
\usepackage{amsmath,amssymb,amsfonts}
\usepackage{algorithmic}
\usepackage{graphicx}
\usepackage{textcomp}
\usepackage[dvipsnames]{xcolor}
\usepackage{enumitem}
\def\BibTeX{{\rm B\kern-.05em{\sc i\kern-.025em b}\kern-.08em
    T\kern-.1667em\lower.7ex\hbox{E}\kern-.125emX}}

\usepackage{tcolorbox}

\usepackage{url}
\usepackage{comment}

\begin{document}

\title{Re-Envisioning Command and Control}

\author{Kaleb McDowell$^{1}$, Ellen Novoseller$^{1}$, Anna Madison$^{1}$, Vinicius G. Goecks$^{1}$, and Christopher Kelshaw$^{2}$ \\ \\
\IEEEauthorblockN{$^1$ Humans in Complex Systems, U.S. DEVCOM Army Research Laboratory, \\ Aberdeen Proving Ground, MD, USA}
\IEEEauthorblockN{$^2$ U.S. Mission Command Battle Lab, Futures Branch, Ft. Leavenworth, KS, USA} \\ 
\IEEEauthorblockN{kaleb.g.mcdowell.civ@army.mil,
ellen.r.novoseller.civ@army.mil, anna.m.madison2.civ@army.mil, \\ vinicius.goecks@gmail.com, christopher.j.kelshaw.civ@army.mil }
\thanks{This research was sponsored by the Army Research Laboratory and was accomplished under Cooperative Agreement Number W911NF-23-2-0072. The views and conclusions contained in this document are those of the authors and should not be interpreted as representing the official policies, either expressed or implied, of the Army Research Laboratory or the U.S. Government. The U.S. Government is authorized to reproduce and distribute reprints for Government purposes notwithstanding any copyright notation herein.}
\thanks{This paper was originally presented at the NATO Science and Technology Organization
Symposium (ICMCIS) organized by the Information Systems Technology (IST) Panel, IST-205-RSY – the ICMCIS, held in Koblenz, Germany, 23-24 April 2024.
}
}

\maketitle

\begin{abstract}
Future warfare will require Command and Control (C2) decision-making to occur in more complex, fast-paced, ill-structured, and demanding conditions. C2 will be further complicated by operational challenges such as Denied, Degraded, Intermittent, and Limited (DDIL) communications and the need to account for many data streams, potentially across multiple domains of operation. Yet, current C2 practices---which stem from the industrial era rather than the emerging intelligence era---are linear and time-consuming. Critically, these approaches may fail to maintain overmatch against adversaries on the future battlefield. To address these challenges, we propose a vision for future C2 based on robust partnerships between humans and artificial intelligence (AI) systems. This future vision is encapsulated in three operational impacts: streamlining the C2 operations process, maintaining unity of effort, and developing adaptive collective knowledge systems. This paper illustrates the envisaged future C2 capabilities, discusses the assumptions that shaped them, and describes how the proposed developments could transform C2 in future warfare. 
\end{abstract}


\section{Introduction}

On the future battlefield, Command and Control (C2)\footnote{The first ``C" in C2, command, is the authoritative act of making decisions and ordering action, with key components being authority, responsibility, decision-making, and leadership. The second ``C" in C2, control, is defined as the act of monitoring and influencing command action through direction, feedback, information, and communications \cite{army2022FM6-0}.} will take place in increasingly complex, dynamic, and challenging situations. Modern, knowledge era technologies are already creating dilemmas for current, industrial era approaches to C2~\cite{Toffler1980thirdwave, payne2021warbot, crilly2022prosecuting}, such as operating with Denied, Degraded, Intermittent, and Limited (DDIL) communications between dispersed or isolated friendly forces. As emerging technologies push the world into what may be considered an intelligence era (i.e., one in which humans partner with artificial learning technologies), creating and maintaining decision advantage---such that our forces make more timely and effective decisions than our adversaries---will require re-envisioning robust C2 systems\footnote{C2 systems specify arrangements of people, processes, networks, and command posts~\cite{army2019ADP6-0}.} and how they should operate. 

A foundational challenge to re-envisioning C2 systems consists of effectively adapting a fundamentally human endeavor\footnote{Referring to the nature of the endeavor as focused on human behavior even if aided by technology.} to include human-machine partnerships. Many modern C2 approaches arose from the era when armies outgrew their communication capabilities, prompting reformers, including 19th-century Prussian field marshal Helmuth von Moltke, to create training and operational techniques, tactics, and procedures (TTPs) that allowed for the army hierarchy to perform C2 on the principles of commander’s intent~\cite{cecil1973helmuth}. We propose here that many of the TTPs in practice today have evolved human-centric features (such as repetition, multi-form communication, group ideation, and reformulating concepts) to ensure commanders and staff under operational stressors and throughout the hierarchy have sufficient situational understanding and trust to effectively maintain unity of effort. The challenge arises in that attempts to integrate advanced technologies often focus on replacing the very human-centric features that are critical to building unity of effort within the command. 

Herein, we envision a future in which the U.S. Army embraces the respective strengths and requirements of human and emerging machine intelligence to achieve enhanced C2 that outperforms decisions made by either humans or artificial intelligence (AI) alone. We will discuss the assumptions that frame the vision and then lay out a prospective C2 system capability in terms of three operational impacts: streamlining the C2 operations process,\footnote{The \textit{C2 operations process} is the Army's framework for organizing and putting C2 into action, and consists of the four major C2 activities performed during operations: planning, preparing, executing, and continuously assessing~\cite{army2019ADP5-0}.} maintaining unity of effort, and developing adaptive collective knowledge systems.

\section{Framing the Vision}

In re-envisioning C2 for a future intelligence era, we make two primary assumptions. As discussed above, we first assume that the future operating environment will present increasingly complex, dynamic and challenging situations. Examples include: future battlefields will experience increased and more effective lethality, pushing the need for distributed C2 systems; effective C2 will require integration across many real-time information streams while operating with DDIL communications between dispersed or isolated friendly forces; and maintaining decision advantage on the battlefield will force the military decision-making process (MDMP)\footnote{Within the planning phase of the operations process, the MDMP defines the series of steps used to produce a plan or order~\cite{army2022FM6-0}. The full MDMP consists of seven steps, from mission receipt to orders production, dissemination, and transition~\cite{army2022FM5-0}.} to be performed at decreasing timescales~\cite{farmer2022four}, for instance, to exploit brief windows of opportunity or adapt to changing risk conditions on the battlefield. 

Our second primary assumption is that technological advancements will fundamentally change the human-technology relationship from that of the current C2 system. For example, technological speed, complexity, and intelligence will enable decisions and actions that outstrip human capabilities under traditional human-machine integration methods; the highest performing emerging AI capabilities will be more akin to ``alien'' intelligence than to animal intelligence, creating fundamental problems for human understanding; and the current rapid pace of technological advancement will be accelerated, resulting in rapid technology fielding and in-field learning by both friendly and enemy forces, which is incompatible with traditional military approaches to Soldier training. 

To achieve decision dominance under the conditions that these assumptions present, most visions of future C2 systems point to the need to integrate both human and machine intelligence. However, the nature of the human-machine integration that frames these concepts can be wildly different as discussed below (see Metcalfe, Perelman et al.~\cite{metcalfe2021systemic} for a detailed discussion of human- versus machine-centric integration). At one extreme, human-centric visions focus on aiding humans to more effectively perform the fundamental C2 tasks (e.g., technologies ranging from human performance-enhancing techniques to individualized, intelligent commander decision support tools). At the other extreme, machine-centric visions focus on leveraging advanced intelligent systems to replace Soldiers in future C2 systems (e.g., replacements may include many of the command staff that perform the fundamental C2 tasks that underlie critical decisions). We forego both extremes and propose that the complexity, dynamics, and challenges of the future operating environment will force the effective integration of significant human resources within C2, while continuous technological advancements will force fundamental shifts in the roles and actions of humans in future C2. We frame the problem as focusing on creating human-machine partnerships~\cite{metcalfe2021systemic} capable of adapting to battlefield changes while inherently enabling unity of effort across large heterogeneous groups of humans and machines. 

\section{Re-envisioning C2}

Recent breakthroughs in interactive machine learning (ML) and hybrid human-machine intelligence 
indicate the potential for humans and AI to work together to leverage their respective strengths~\cite{case2018become, scharre2016centaur, zhang2019leveraging}.
There is, however, no one-size-fits-all approach to human-machine integration. Metcalfe, Perelman et al.~\cite{metcalfe2021systemic} introduce a landscape of human-AI partnership, which posits that the nature of human-AI interaction depends on task complexity, decision timescales, and information certainty. While some simpler sub-tasks or functions may best be solved by technologies alone, many of the fundamental tasks that underlie effective future C2 fall within the complex landscape requiring effective human-machine partnership. Furthermore, over time, more and more sub-tasks and functions are expected to be solved via technology, which in turn will shift the nature of the interactions on the complex landscape. These factors make it difficult or even inappropriate to specify a type of future C2 human-machine partnership. To overcome this challenge, we take a multi-pronged approach. Below, we discuss how future C2 systems can potentially impact three specific aspects affecting the future C2 operations process. We augment each of these discussions with a future Soldier ``account'' aimed at inspiring imagination. Finally, we augment this discussion with two companion papers~\cite{madison2024scalable,goecks2024coagpt}. The first is a research paper describing a large language model-driven course of action (COA) generation capability (see Goecks and Waytowich~\cite{goecks2024coagpt}) that illustrates the near-term capability at the heart of the examples herein, and the second outlines a broad research program illustrating the types of science and technology that potentially will enable the discussed technologies herein (see Madison, Novoseller, Goecks, et al.~\cite{madison2024scalable}). 

\begin{tcolorbox}[colback=blue!5!white, float, floatplacement=t, parbox=false]
\section*{Staff Sergeant (SSG) Hill}

My first military occupational specialty in the Army was data wrangling. I examined battlefield data, e.g., images and audio, and compared them to the machine’s interpretations. At first it was simple stuff like targets, but over time the machines got better, and later I looked at whole phases of an operation for concepts like danger. To this day, I still data wrangle for short ``sprints'' when major events occur. 

For my next deployment, I was assigned to an intelligence functional cell. In this role, I would analyze potential threat and friendly COAs using specialty AI tools that would advise me on potential intelligence-specific risks and benefits. The tools also allowed me to ask questions to help me understand the rationale behind the risk and benefit analyses as well as allowing me to compare alternatives. This was a pretty intense job, since new potential COAs were regularly being generated as new information arose. Our goal was to be familiar enough with viable possibilities that we could clearly articulate an intelligence recommendation on any COA that we were asked about. Following intelligence, I spent time in fires and sustainment cell analysis teams. Through these three roles, I felt I had a pretty good idea of how to analyze a COA.

I am now on a COA development team within the Future Operations integrating cell, and it’s a way more dynamic process! The machines are great at generating and adjusting COAs as new information arrives and is verified. I have had several roles on this team. I first focused on ensuring the COAs were aligned to mission intent and parameters. In another role, I was part of a team that would identify which COA options are potentially viable, and I liaised with the analysis teams to evaluate these options and generate a cohesive assessment. Right now, I perform COA adjustments. In this role, I have many tools to create options, including: speaking to a plan generator, sketching on a board, moving assets in a virtual sandbox, and/or just highlighting part of a plan and saying ``good,'' ``bad,'' ``fix that.'' In each case, the machines immediately adjust the COA and provide feedback on risks and benefits, and together we quickly iterate until we reach a solution.
\end{tcolorbox}

\subsection{Streamlining the C2 Operations Process}

We anticipate that human-AI interaction will streamline the C2 operations process (i.e., combining multiple steps in a process or executing processes simultaneously) in contrast to the current C2 approach, which is more linear and consists of many sequential steps~\cite{army2022FM5-0, army2022FM6-0}. One of the primary outcomes of streamlining C2 is to dramatically reduce the timelines of deliberate planning by enabling simultaneous COA development and analysis, as well as optimizing elements of control through rapidly generating and selecting branches and sequels. \footnote{Branches and sequels adjust plans and orders beyond the initial stages of an operation. Branches are ``contingency options built into the base plan used for changing the mission, orientation, or direction of movement of a force\textellipsis based on anticipated events, opportunities, or disruptions caused by enemy actions and reactions''~\cite{army2022FM5-0}. A sequel is ``the subsequent operation or phase based on the possible outcomes of the current
operation or phase''~\cite{army2022FM5-0}, for instance a counteroffensive following a defense.}
Such streamlining has the potential to bring greater rigor to C2 processes, e.g., increasing analytic power with fewer available personnel, 
which could create decision advantage when the need for hasty, short duration planning arises (e.g., Crisis-Action Planning).

At its core, we envision an interactive technological capability, or Intelligent Course of Action Suite (iCOAs), that can almost instantly integrate available information into a high-resolution COA that contains much more detailed information compared to current COAs. Such a capability includes multiple intelligent technologies and processes that simultaneously create, analyze, and compare plans along an array of mission and operational variables and perspectives critical to operations (e.g., fires, maneuver, and sustainment warfighting functions; moral, ethical, sociocultural, and political implications; adversarial, civilian, and friendly actions; etc.). Further, it is expected that the iCOAs runs in real-time to process data from battlefield sensors and sources to maintain up-to-date situational awareness and generate branches and sequels as needed. Early instantiations of the suite are likely to heavily rely on humans to refine, analyze, and compare the plans; however, as the technologies evolve, we foresee a highly integrated human-machine partnership with both humans and machines filling multiple roles in C2 (see SSG Hill). 

This vision builds on ideas proposed by Farmer~\cite{farmer2022four}, which applies calculations typically performed during COA analysis within the COA development step to rapidly compare and optimize over many COA alternatives in an example scenario. In contrast, we anticipate an iCOAs that not only leverages the search capabilities of AI, but also draws on the combined strength of humans and AI to maximize creativity, exploration, military expertise, and common sense, as well as address ethical dilemmas, across the C2 operations process. The envisioned streamlined iCOAs will present and explain proposed COAs to personnel in an intuitive format, for instance via modalities such as language, annotated maps, or trees of possible outcomes, while providing standard COA analysis outputs such as synchronization matrices~\cite{army2022ATP5-0.2-1} and metrics such as success probabilities and uncertainty estimates that account for environmental variables and potential threat COAs. C2 personnel will input data to the system as it becomes available; analyze presented COA alternatives and associated data; interactively adapt AI-generated COAs based on domain expertise, common sense, and changing situational needs; and make final COA selections. 

C2 personnel will dynamically interact with the AI system to adapt proposed COAs in a variety of ways depending on situational needs. C2 personnel could describe objectives (e.g., ``Take control of the enemy asset''), issue corrections as needed to adapt and fine-tune proposed plans (“approach from the opposite direction to avoid being seen”), and specify constraints that prohibit the AI from choosing undesirable actions (e.g., ``maintain a distance between 800 to 1000 m above mountain terrain''). When unanticipated changes occur, personnel could relay feedback to the AI system to indicate how to adapt the plan to emergent risks and opportunities (e.g., ``the enemy has weaponized trees, so avoid trees when possible''). C2 personnel could adjust COA selection criteria to trade-off between multiple objectives, for instance upweighting the importance of controlling an enemy asset or penalizing actions that exacerbate danger to personnel or equipment, and could give feedback on multiple COA alternatives to teach the system which COA features are preferable. It is also envisioned that C2 personnel and the iCOAs each have the capability to ask for help from the other on decisions, tasks, or challenges as needed.  Ultimately, a COA would be selected once the commander is satisfied with the presented alternatives; this COA could continue to be updated at multiple decision points and as battlefield conditions evolve. 

The envisioned capability could allow for compression of battle rhythms\footnote{\textit{Battle rhythms} are a deliberate daily cycle of command, staff, and unit activities established by commanders, consisting of a series of meetings, briefings, and other activities synchronized by time and purpose~\cite{army2022FM6-0}.} though more sophisticated data processing pipelines, leading to faster actionable intelligence for more efficient decision-making. The high-resolution nature of iCOAs would allow for faster adoption of COAs down the chain of command. Systems that enable such streamlining would accelerate C2 operations and potentially change the physical footprint of the Army on the battlefield by allowing for organizational changes in the Army, changing  personnel requirements, and/or altering the required resources for logistics and sustainment. In addition to deliberate planning, the real-time nature of the capabilities allows for the same capabilities to be used during both initial planning and dynamic operations. Running continuously, the iCOAs would update predicted outcomes of COAs based on battlefield progress and generate potential adjustments as unanticipated events and information emerge. These capabilities would enable decision-making at fast enough speeds to enable forces to exploit or create brief windows of opportunity and quickly adapt to dynamic battlefield conditions to gain decision advantage. 

\subsection{Maintaining Unity of Effort}

\begin{tcolorbox}[colback=blue!5!white, float, floatplacement=t!, parbox=false]
\section*{Major Mahomes}

``Alert, alert, communications are actively being jammed!'' We knew they would jam us, but my iCOAs-S is indicating it's occurring much earlier than expected. We had not anticipated adversary activity until we had at least reached phase line PL Coppermine. Suddenly, a muted boom, and we can see the skyline light up with rocket trails and flashes of explosions---it looks like they are trying to block the entire movement corridor. 

Using on-board sensors, our COP\footnote{Common Operating Picture} is indicating that the attack is targeting an area approximately six klicks to the north-north-east and covers an area at least 20 klicks wide. It is also indicating that there is potentially active fire over 30 klicks to the east, south-east, and south. I focus on the iCOAs-S: it indicates that our current COA is suboptimal to meet our mission objectives and has three alternatives prioritized by expected success rate. The first two COAs have a much higher success rate, and so I focus on analyzing those. While I am doing my analysis, I ask iCOAs-S what’s the primary difference between the COAs. It responds that COA Alpha is a coordinated northern-focused attack that is more conservative, will only partially attain mission goals, and has a slightly higher expected casualty rate. It is highly likely that all units in this operation will have COA Alpha as a top alternative, and within the next 5-10 minutes via on-board sensors, units will be able to confirm high unity of effort. 

COA Bravo is a multi-pronged attack requiring northern, eastern, and southern movements. If successful, it will fully attain mission goals and has a slightly lower expected casualty rate. However, it is unlikely that the 773th Battalion operating far to the west would have COA Bravo as an alternative given the predicted range of the jamming, terrain, weather, and the capabilities of their on-board sensory. I ask iCOAs-S for a plan to get the critical information to the 773th Battalion. ``Update'' flashes on the screen, on-board sensors indicate that Captain Nacua over in Zulu Company already taken action to transmit emergency information to the 773th. I don’t fully understand how the technology works, but it can get very small amounts of critical information through jamming at some increased risk to the unit. I can see that COA Bravo’s expected success rate is increasing rapidly. I quickly work with iCOAs-S to refine my unit’s specific COA and order our unit to starting executing Bravo. For the next 5 minutes, I monitor the unity of effort meter and the emergency communications channel, and it seems everyone is coordinated.
\end{tcolorbox}

As the Industrial Revolution ballooned fielded armies to over 1 million personnel, unity of effort became a critical barrier to effective operations. In the 19th century, Helmuth von Moltke overcame this barrier through a highly selective approach to developing interchangeable officers and organizations who could effectively predict each other’s actions without direct communications~\cite{cecil1973helmuth}. Akin to the Prussian’s approach, we anticipate that human-AI systems will be able to enhance the ability of unit leaders across echelons to understand and predict each other’s actions under future operating environment conditions such as DDIL (see Major Mahomes for a far future example). 

One of the primary outcomes of these human-machine systems is faster coordination with lower direct communication requirements. These critical capabilities will also provide lower echelon leaders a far greater capability to consider the broad scope of battlefield operations in their decision-making abilities, thus enabling more complex and coordinated action. 

We envision an advanced scalable version of iCOAs (iCOAs-S) that can develop nested COAs across multiple echelons, such that C2 system personnel will interact with the suite at multiple hierarchical levels. For instance, at higher echelons in the planning phase, commanders and multiple staff teams might work with the iCOAs-S to rigorously develop and analyze hundreds of friendly and enemy battlefield COAs, while at lower echelons during operations, the iCOAs-S may serve more akin to a decision recommender to individual commanders or unit leaders as battlefield conditions unfold. 

Critically, we envision that iCOAs-S would maintain COAs across multiple hierarchical levels while integrating information from each echelon. Running locally at each echelon, this ``high-fidelity'' resource would be 
able to include enemy and friendly anticipated actions (including considering the biases of their respective commanders) during unanticipated operational events, but with a broader inclusion of the anticipated actions held by different perspectives at each echelon (i.e., operational levels through to tactical levels).
We assume that given DDIL, locally run instances of iCOAs-S would not all have the same real-time information; however, iCOAs-S would be able to project what real-time information is likely to be known by other individual units and predict behaviors accordingly (see Major Mahomes 773th Battalion example). We propose that this enhanced predictability (tailored to the landscape of human-machine partnership) inherently allows units to maintain unity of effort, and has the potential to dramatically accelerate coordinated action and reduce communication requirements. 

A second primary outcome is the ability of iCOAs-S to enable C2 system personnel to more effectively and rapidly understand how potential COAs lead to outcomes. One of the core capabilities of the iCOAs envisioned above is the near real-time analysis of COAs. We foresee iCOAs interfaces that allow for commanders and staff to work step-by-step through the COA development process. However, unlike current processes where COA analysis comes after COA development is completed, we expect iCOAs to provide running estimates of key mission outcome metrics as each option in the decision process is weighed. That is, iCOAs will be able to evaluate even high-level comparisons (e.g., frontal attack vs. envelopment), generating and analyzing potential COAs in near real-time. In early implementations of the iCOAs, it would be expected that commanders and staff pursue each step of the COA development process and use iCOAs as a decision aid, but as the suite is developed over time, personnel will learn in which steps the suite performs very well and in which steps human interaction is critical. In its end state, we can envision the process changing from linear individual COA development to a parallelized exploration in which C2 personnel rapidly identify key decisions across the entirety of the development and analysis process, enabling iCOAs to collaboratively generate multiple viable COAs simultaneously. We propose that the type of explorative decision-outcome linked system described here will rapidly build the common understanding in commanders and staff that is more critical to unity of effort than some of the historically human-centric C2 system features discussed above. 

\subsection{Developing Adaptive Collective Knowledge Systems}

\begin{tcolorbox}[colback=blue!5!white, float, floatplacement=t, parbox=false]
\section*{Lieutenant Colonel (LTC) Mostert}

The mission was successful, but there were a few things that could have gone better. We are back at Forward Operating Base Justice, and as part of an after-action-review, I am focusing my team on answering the question, ``Why didn’t we expect enemy action before we reached PL Coppermine?''

We downloaded data from all the units and sensors that were deployed on the battlefield and are stepping through a recreation of the events in the 5 minutes surrounding the onset of jamming. Our iCOAs-SA technologies are indicating inconsistencies between projected enemy capabilities and actual events. Specifically, it is saying that the enemy doesn’t have the capability to jam our systems given their locations at the time of the attack! How did this happen?

I set my teams on two different courses of action. Firstly, since the jamming did occur, we need to have that knowledge baked into future plans. The team works with the intelligence analysis team to pass on the specific capabilities observed. Intelligence will look through databases of previous missions to identify any examples of this capability. Then, they will update both human and iCOAs-SA resources with their new estimates of enemy capabilities. 

Second, I have another unit looking for anomalies on the battlefield that deviate from the intelligence we had prior to the mission. We have terabytes of data across all sensor streams, so the team employs multiple AI algorithms to look for different types of anomalies. Once anomalies are detected by the AI, the team pulls in data wranglers so that human analysts have ``eyes-on'' the isolated data. While this process eliminates many of the potential anomalies, one particular anomaly is promising. About 2 minutes before the jamming started, the human-AI teams identified three abandoned automobiles emitting a low-frequency radio signal. The team then searched for this combination across the battlefield and found six other examples of this particular combination. 

I immediately notify HQ and begin planning a mission to assess this potential new technology.
 
\end{tcolorbox}

Our assumptions in framing the vision underscore the need for systems and technologies to be adaptive. We assume that future C2 operations will face more dynamic environments, faster decision cycles, more rapid technology fielding---and once fielded---more rapid technology updates and in-field learning, with increased lethality. These assumptions have ramifications for both materiel and non-materiel aspects of future C2 systems, further pointing to the need for effective partnership between humans and technology. 

We envision an adaptive version of iCOAs-S, iCOAs-SA, that leverages advances in interactive machine learning to integrate human feedback and large databases of shared collective knowledge into decision-making. Shared collective knowledge may include experience from past missions, previous COAs, after action reviews (AARs), and other relevant data, and will constantly expand as new experience is gained. The suite will be able to learn from virtually every interaction it has with humans; however, not all of these interactions will lead to effective, stable behaviors. To overcome this challenge, we envision a fundamental change in the nature of iCOAs-SA's human-machine interactions compared to that of current C2 systems. 

As introduced in the SSG Hill data wrangler role and expanded in the LTC Mostert AAR example, C2 system personnel will be responsible for guiding and teaching the iCOAs-SA to effectively adapt throughout C2. We believe that this guidance will be necessary for two primary reasons: first, the nature of the C2 is so complex and human-centric (e.g., see ethical, moral, legal, political considerations) that humans will need to provide inputs for the systems to learn from (e.g., demonstrations, decisions, feedback) across the multiple steps of the C2 operations process as well as within different stages of data processing, integration, and interpretation. As discussed before, as the iCOAs-SA adapts, the exact nature and roles of these C2 personnel may shift; however, we propose that changing environments and mission requirements drive the need for continued human involvement across steps and stages. Further, we assume that when iCOS-SA is introduced in novel environments or missions, the need for human guidance will be ramped up to identify and mitigate potentially ineffective behaviors before subsiding as the system learns. Second, humans will be critical to control mechanisms, such as human-machine AARs emplaced by trained leaders (see LTC Mostert), that modulate at what times and from which inputs the suite adapts to ensure effective, ethical, and moral behaviors over time. We expect that the changing nature of the human-machine interactions will be significant enough to influence how future Soldiers are recruited, assigned positions, and trained.  To be effective, human-guided iCOAs-SA will have to be sensitive to differences between various interacting humans (e.g., different preferences or biases), as well as sensitive to human dynamics (e.g., state changes that may influence biases or errors generally).

In addition to enabling adaptation, the iCOAs-SA potentially has another critical capability. As it is built on shared collective knowledge, over time it will inherently capture the knowledge and expertise that is lost in current Soldier rotations. Primarily, this ensures that the lessons iCOAs-SA learns over time are not lost when future C2 system personnel rotate. Secondarily, this capturing of knowledge may help reduce some of the burden associated with personnel rotations, making the overall system more efficient. Finally, the adaptability within the suite may also enable more effective tailoring of information to C2 system personnel based on their roles, functions, and situational context. 

\section{Conclusions}

Current C2 systems, which have evolved as a fundamentally human endeavor, will be challenged as emerging technologies push the world into the intelligence era. However, transitioning to an intelligent, technology-driven approach to C2 has to-date been difficult. We argue that the core challenge of C2 for the future operating environment, involving coordination of large groups of people and machines, requires an approach that emphasizes the integration of both technological capabilities and critical aspects of human teaming such as maintaining unity of effort. 

We outline a human-machine partnership-focused approach to future C2 systems using the notional example of an adaptive, scalable, intelligent COA development and analysis suite of technologies, iCOAs-SA. Through providing examples of how such a suite of technologies could impact operations in three ways---streamlining the C2 operations process, enabling unity of effort, and creating effective adaptability---we illustrated a future vision of C2 that addresses assumed future operational challenges while also balancing materiel and non-materiel considerations. 

While our approach largely focuses on providing examples of a C2 system possible in the far future, we posit that by focusing on human-machine partnerships as a core principle of system development, intelligent systems could be designed that intentionally evolve over time. That is, initially, suites of technologies could operate as aids in a largely human-dominated system akin to current C2, while over time with experience and refinement, the nature of the interactions could evolve to the proposed partnerships described herein. Consistent with this, there are several recently-developed AI tools for the MDMP focused on accelerating  mission analysis, such as understanding commander's intent~\cite{schadd2022machine}, and  COA development and analysis~\cite{yuksek2023intelligent,goecks2024coagpt,schwartz2020ai}. Further, we acknowledge that while we frame the problem in terms of human-machine partnerships, progress in human- and machine-centric approaches will be necessary and effective when integrated with the types of advancements envisioned herein, particularly with respect to the technology and infrastructure available at the tactical edge. For instance, mirroring industry trends of integrating AI into devices such as smartphones and personal computers, we predict that the computing and power requirements for future AI technologies will be met for warfighters at the tactical edge.

\section*{Acknowledgements}
The authors would like to thank Benjamin T. Files, Nicholas Waytowich, Alfred Yu, Vernon Lawhern, Steven Thurman, Howie Brewington, and Jay Goodwin for engaging in helpful discussions.

\bibliographystyle{ieeetr}
\bibliography{References}

\end{document}